\documentclass{article}
\usepackage{waspaa23,amsmath,graphicx,url,times}
\usepackage{color}
\usepackage{algorithm}
\usepackage{algorithmic}
\usepackage{amsfonts}
\usepackage{cite}
\usepackage{subfig}
\usepackage{setspace} 
\usepackage{caption}
\usepackage{dashrule}
\usepackage{ifthen}
\title{Hybrid noise shaping for audio coding using perfectly overlapped window}
\name{Byeongho Jo$^{1}$\sthanks{This work was supported by Electronics and Telecommunications Research Institute (ETRI) grant funded by the Korean government [23ZH1200, The research of the basic media/contents technologies].}, Seungkwon Beack$^{2}$}
\address{$^{1,2}$ Electronics and Telecommunications Research Institute, South Korea}

\begin{document}

\ninept
\maketitle

\begin{sloppy}

\begin{abstract}
In recent years, audio coding technology has been standardized based on several frameworks that incorporate linear predictive coding (LPC). However, coding the transient signal using frequency-domain LP residual signals remains a challenge. To address this, temporal noise shaping (TNS) can be adapted, although it cannot be effectively operated since the estimated temporal envelope in the modified discrete cosine transform (MDCT) domain is accompanied by the time-domain aliasing (TDA) terms. In this study, we propose the modulated complex lapped transform-based coding framework integrated with transform coded excitation (TCX) and complex LPC-based TNS (CTNS). Our approach uses a 50\% overlap window and switching scheme for the CTNS to improve the coding efficiency. Additionally, an adaptive calculation of the target bits for the sub-bands using the frequency envelope information based on the quantized LPC coefficients is proposed. To minimize the quantization mismatch between both modes, an integrated quantization for real and complex values and a TDA augmentation method that compensates for the artificially generated TDA components during switching operations are proposed. The proposed coding framework shows a superior performance in both objective metrics and subjective listening tests, thereby demonstrating its low bit-rate audio coding.
\end{abstract}


\section{Introduction}
\label{sec:intro}

Audio coding technology has evolved to incorporate transform-based codecs and linear prediction coding (LPC)-based speech codecs \cite{AMRWBp,usac,evs,mpegh}. The modified discrete cosine transform (MDCT) has been used as the main transformation in numerous audio codecs owing to its ability to provide critical sampling and perfect reconstruction\cite{MPEG1,AAC,bosi2002introduction}. The speech coding methods based on the LP analysis have been developed to efficiently compress the residual signal obtained using the all-pole filtering with LPC coefficients in the time and frequency domains \cite{gibson2005speech}. Particularly, the transform-coded excitation (TCX) technology adopted in AMR-WB+\cite{AMRWBp} has been developed and advanced with numerous techniques to quantize and code the LP residual signal in the frequency domain. 

The original TCX was proposed to quantize the magnitudes and phases of the complex discrete Fourier transform (DFT) coefficients separately after LP filtering and removing the pitch component \cite{lefebvre1994high}. However, it was later modified to quantize the DFT coefficients by stacking the real and imaginary parts using algebraic vector quantization (AVQ) and extended AVQ \cite{xie1996embedded,ragot2004low}. Fuchs {\textit{et al.}} proposed an MDCT-based TCX framework and demonstrated a better coding performance of the scalar quantization by quantizing the MDCT coefficients instead of the DFT coefficients\cite{fuchs2009mdct}. In their later work, they proposed a framework that selectively operated temporal noise shaping (TNS) before the conventional MDCT-based frequency-domain noise shaping (FDNS) \cite{fuchs2015low}. 

The TNS was first proposed to address pre-echo artifact problems when performing transform-based coding with large-sized temporal frames for transient signals, which has been adopted by a series of audio coding standards. Vaalgamaa {\textit{et al.}} proposed a framework that selectively operates the TNS on the warped LP filtered residuals \cite{vaalgamaa1999audio}, which reverses the order of the TNS from the coding framework in \cite{fuchs2015low}. These frameworks with TNS operate TNS on the MDCT coefficients, thereby resulting in time-domain aliasing (TDA) artifacts because they predict and shape the envelope of the time domain containing the TDA components. Attempts to address this problem have been made by using windows with less overlap or by window switching \cite{herre1996enhancing,allamanche1999mpeg,schnell2007enhanced}. However, it does not address the problem. Here, it is possible to operate a TNS tool with perfect duality using DFT. Jo {\textit{et al.}} recently published a study on this \cite{jo2023audio}, wherein they named the complex LPC-based TNS (CTNS). However, since doubling of the data rate cannot be avoided when using DFT, using a low overlap window seems like a compromise. 

In this study, we propose an audio coding framework that not only retains the gains obtained by combining DFT and TNS but also increases the frequency resolution by using a perfectly overlapped window. Instead of using DFT, the proposed framework incorporates the modulated complex lapped transform (MCLT), a complex form of MDCT, for the transform to achieve TDA artifact-free TNS by maintaining the duality of the complex-valued coefficients and temporal envelope. Additionally, the coding performance for tonal signals can be enhanced by using a perfect 50\% window. To achieve perceptually seamless transitions in the mode transitions, a unified quantizer for complex and real input values and an algorithm to arbitrarily augment the TDA component in the MCLT frame are designed. Finally, we propose an algorithm that predicts the frequency envelope from the quantized LPC coefficients and uses it to adaptively allocate the required target bits per time frame and sub-band to calculate the sub-band scale factors. Here, the scale-factor determination procedure is complemented. 

\section{TNS and window shapes}
\label{sec:ctns}
\begin{figure}[t]
\begin{minipage}[b]{1.0\linewidth}
  \centering
  \centerline{\includegraphics[width=8cm]{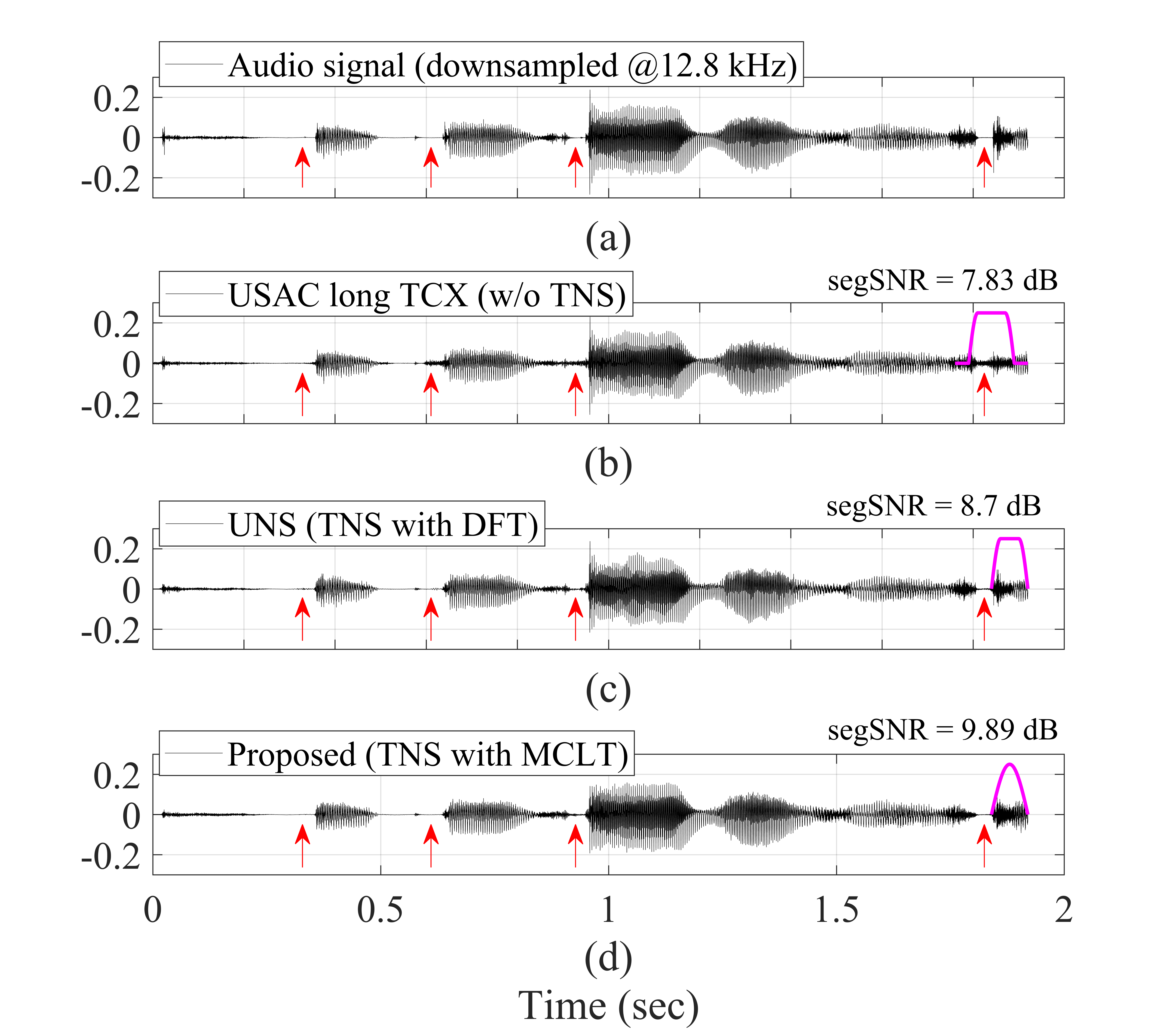}}
\end{minipage}
\caption{TNS effects with different transforms and windows. (a) downsampled audio signal (es01). The reconstructed signal of (b) USAC long TCX (without TNS), (c) UNS (CTNS with DFT), and (d) the proposed system (CTNS with MCLT). All reconstructed signals are coded at $12$ kbps. Red arrows indicate the temporal positions where the TNS tool operates effectively. The window function for each method is depicted as the magenta solid line.}
\label{fig:effect}
\vspace*{-3mm}
\end{figure}

In early audio coding research, the problem of pre-echo arose owing to the use of large-size temporal frames. To tackle this, TNS was proposed \cite{herre1996enhancing}. It has been successfully adopted and used in numerous audio coding standards \cite{AAC,usac}. TNS is a technique for estimating frequency coefficients based on the duality of the time/frequency analysis and accordingly shapes the envelope in the time domain. However, when operating in the MDCT domain (commonly used for audio coding), it incorrectly predicts the temporal envelope of frames containing TDA components \cite{liu2008compression}. To address this problem, the original TNS minimized the unwanted distortion of the TNS by using a small overlap window. However, the small overlap window has properties close to a rectangular window, thereby resulting in a deteriorated frequency interpretation \cite{bosi2002introduction}. This can be solved mainly by operating the TNS in the DFT domain as discussed in the recent study by Jo {\textit{et al.}}\cite{jo2023audio}. However, its disadvantage includes the critical sampling property of the MDCT that cannot be exploited. Although TNS artifacts due to the TDA can be avoided by CTNS, DFT doubles the information to be quantized, which again forces the use of small overlapping windows to increase the frame rate to compensate. However, the use of such a small overlap window has disadvantages including deteriorating frequency resolution, thereby weakening the coding performance for the tonal signals. Rather than using small overlap windows, we propose a framework that uses a perfect 50\% overlap window. To ensure the coding performance of the tonal signals while offsetting the loss of the coding gain owing to the increased frame rate, we propose a switching technique with an integrated transform based on the MCLT rather than using only the DFT.

Figure \ref{fig:effect} illustrates the effectiveness of the proposed framework. Figure \ref{fig:effect}a shows the waveform of an audio signal downsampled to $12.8$ kHz. The reconstructed time signal is shown in Fig. \ref{fig:effect}b with the long TCX mode of the MPEG USAC \cite{usac} and without the TNS using a small overlap window. Because the long TCX uses a large frame and an analysis/synthesis window with degraded frequency resolutions, it can be observed that the pre-echo problem occurs in the region near the onset of the transient signal, represented by the red arrows in the figure. Contrarily, by using the TNS with the DFT proposed in \cite{jo2023audio}, the pre-echoes in those regions are significantly reduced despite using an imperfect window (Fig. \ref{fig:effect}c). Because the proposed framework uses a perfect 50\% overlap window and operates the TNS with the MCLT, the pre-echo is significantly reduced to a similar extent, as shown \ref{fig:effect}c, while increasing the coding gain through better frequency resolution. The segmental signal-to-noise ratio (segSNR) values of the three methods illustrate that the coding performance of the proposed system is improved at a similar bit rate, while maintaining the effectiveness of the MCLT-based TNS despite using a 50\% overlap window. In the next section, the details of the proposed system that utilizes the 50\% overlap window and MCLT-based CTNS are presented.

\section{Proposed framework}
\label{sec:prop}
\begin{figure}[t]
\begin{minipage}[b]{1.0\linewidth}
  \centering
  \centerline{\includegraphics[width=8cm]{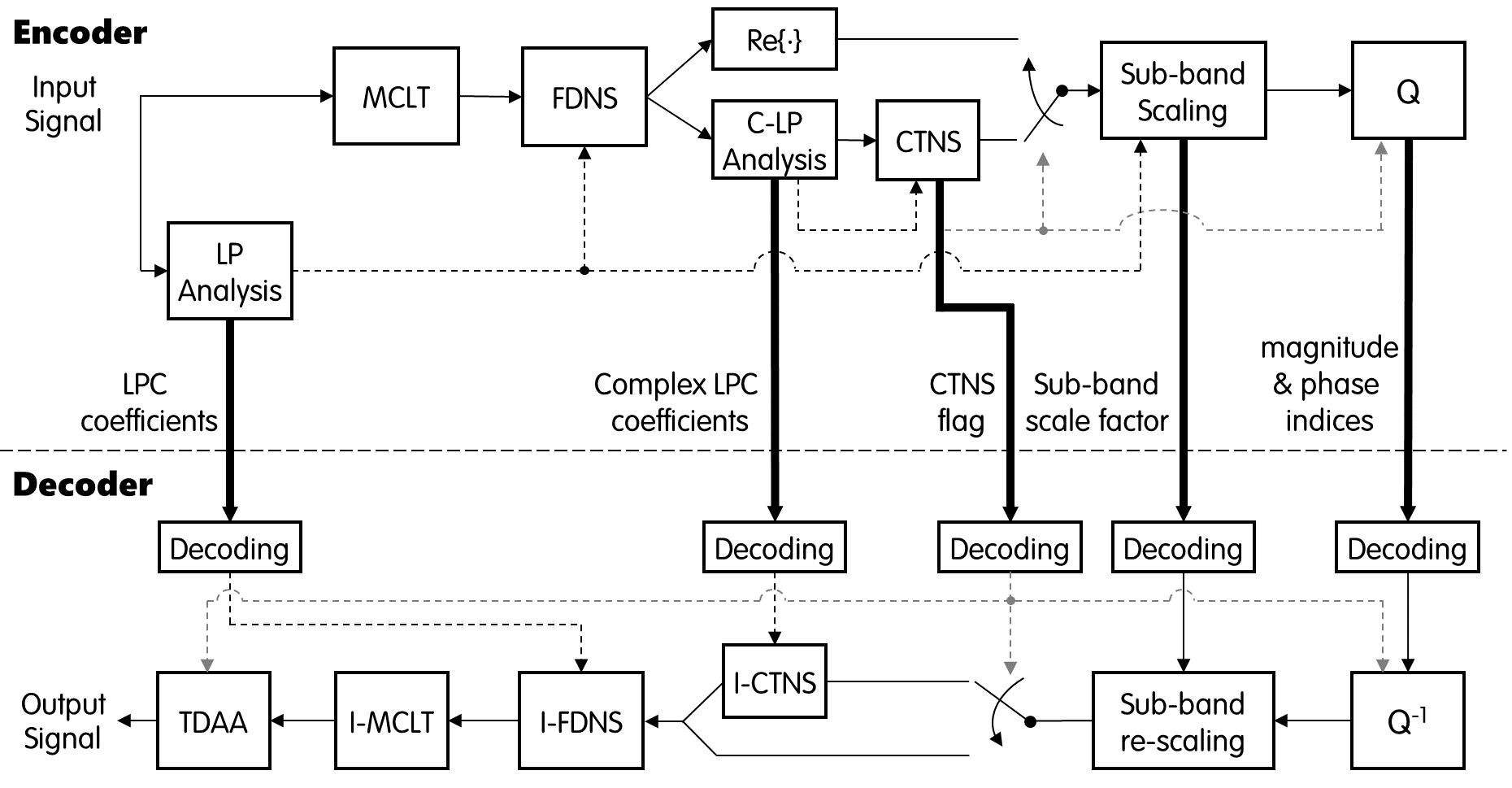}}
\end{minipage}
\caption{Block diagram of the encoder and decoder. Thin and solid lines indicate the signal flows. Dotted lines indicate control signals. The encoded signals are described as thick and solid lines.}
\label{fig:block}
\vspace*{-3mm}
\end{figure}

\subsection{Overview}
\label{ssec:over}
The proposed audio coding framework is shown in Fig. \ref{fig:block}. The encoding part is illustrated at the top side of Fig. \ref{fig:block}. Here, the input time-domain signal is fed into the encoder. The LP analysis is conducted using a hamming-windowed block, which estimates the LPC coefficients for each block. Simultaneously, the input block is windowed and transformed using MCLT \cite{malvar1992signal,malvar1999modulated}, and magnitudes of their coefficients are shaped by FDNS using the quantized LPC coefficients. The FDNS is conducted by subtracting the magnitudes from the frequency envelope estimated by the quantized LPC coefficients. Further, the complex-valued FDNS residuals are fed into the complex LP analysis. The complex LP analysis estimates the complex LPC coefficients by using the Levinson-Durbin algorithm. The estimated complex LPC coefficients are quantized and utilized for the CTNS. Particularly, the FDNS residual coefficients are filtered by the complex all-pole filter with quantized complex LPC coefficients as its filter coefficients, and the prediction gain is calculated for the use of CTNS. The prediction gain is calculated similarly in \cite{jo2023audio}. The CTNS is operated only when the prediction gain is larger than the pre-defined threshold. In this study, the gain threshold for the CTNS is set to $-3$ dB. The CTNS filtering is operated only above $312$ Hz. When the CTNS is deactivated, the real part of the complex-valued FDNS residuals are fed into the sub-band scaling. Here, one bit is allocated for the CTNS flag which indicates whether the CTNS tool is operated and transmitted to the decoder. 

Instead of using a single scale-factor, the proposed system adopts the multiple scale-factors for $8$ sub-bands. The boundaries for the sub-band scaling is modified based on the equivalent-rectangular bandwidth (ERB) scale, which are [$40$, $90$, $140$, $200$, $260$, $330$, $410$, $512$] Hz. The scale-factors are calculated using the allocated bits and power of the tuple of coefficients in each sub-band. The allocated bits are controlled adaptively, and the details for this are elaborated in Section \ref{ssec:bits}. By using the scale-factors to scale, the scaled coefficients are obtained. These coefficients can either be the real or complex values. When the CTNS is deactivated, the scaled coefficients are the real-valued, and when they are not, the scaled coefficients are the complex-valued. When reconstructing a time signal wherein the CTNS is onset or offset, quantization errors can be superimposed and cause undesired coding artifacts. These artifacts can be categorized into two parts including the TDA terms from the MDCT-frame and magnitude discontinuities caused by using different quantization modules. To alleviate the former parts, the time-domain aliasing augmentation (TDAA) is adopted, which is discussed in Section \ref{ssec:TDAA}. 

When different quantizers are used for real and complex values, the latter problem may not be negligible. This can easily be resolved by using a unified quantizer with similar quantization error patterns. Particularly, when the input $A$ is fed into the quantizer, the magnitude is first quantized using a non-uniform quantizer regardless of the types as $I_m=\bigl\lfloor A^{\mathbf p(b)}+0.48\bigr\rfloor$ wherein $\mathbf{p}(b)$ denotes the pre-defined power-factor of $b$-th sub-band. The power-factor vector is set to $\mathbf{p} = [4/3, 4/3, 4/3, 1, 1, 3/4, 3/4, 3/4]$. The implied reason for such a design is to increase the quantization performance for high input magnitudes in the low frequency sub-band and low input magnitudes in the high frequency sub-band. After quantization for the magnitudes, the phases or the signs are quantized based on the CTNS flag. The phases of the complex coefficients are uniformly quantized to $6$ bits when CTNS is on, while the signs of the real numbers are uniformly quantized to $1$ bit when it is off. It is noteworthy that bits for phases can be further reduced by using different unrestricted polar quantization algorithms \cite{vafin2005entropy} as in \cite{yoon2008coding,Jo2022modified}. Here, no bits are allocated for both the phase and sign with zero magnitudes. The magnitude discontinuities can be minimized by using the unified quantization. The quantization indices of the magnitudes and phases or signs are entropy coded and transmitted to the decoder side.

The decoder of the proposed framework is shown at the lower side of Fig. \ref{fig:block}. The decoding operation is conducted by the reverse ways of the encoder. Particularly, the inverse quantization is conducted, and the de-quantized coefficients are re-scaled by the decoded scale-factors. Further, the inverse filtering by the CTNS is selectively operated based on the decoded CTNS flag. The inverse operation of the FDNS and inverse transformation is conducted in turn. The TDAA is selectively conducted based on the CTNS flags of the current and previous time frames. The procedure details of TDAA is elaborated in section \ref{ssec:TDAA}.

\subsection{Adaptive bit allocation using frequency envelope}
\label{ssec:bits}
For the sub-band scaling, the scale-factor for each of them must be determined effectively. The scale-factors are calculated by using modified \ttfamily{SQ\_gain}\normalfont{} function in the MPEG-USAC \cite{usac}. The original function \ttfamily{SQ\_gain}\normalfont{} estimates the global scale-factor based on the energies of the quadruple coefficients and target bits. The target bits are adaptively determined but their adaptation can be done for only a single frame since it utilizes the single global scale-factor. In contrast, in this study, we propose an algorithm to compute scale factors in multiple sub-bands and adaptively determine the target bit for each sub-band. The proposed algorithm uses the frequency envelope (FE) information computed by inverse transformation of the quantized LPC coefficients to adaptively determine the final target bit per frame and sub-band. First, let the computed FE be $H_{dB}(\nu,f)$ in the decibel scale. where $\nu$ denotes the frame index and $f$ the frequency index. Then, one can calculate the FE ratio (FER) by finding the maximum value of the frequency envelope within a sub-band and normalizing it to the sum of all the sub-bands. The corresponding equations are elaborated in \cite{jo2023audio}. Additionally, to consider the energy of each frame, a frame gain is calculated as expressed below: 
\begin{equation}
G(\nu) = 20\log_{10}{\{\textstyle\sum_{f} {H(\nu,f)}\}/N},
\end{equation}
where $H(\nu,f)$ is the FE in the linear scale and $N$ is the number of coefficients. Using the frame gain, FER, predefined fixed and additional bits, and FER threshold as input, the target bits are adaptively calculated and fed into the modified \ttfamily{SQ\_gain}\normalfont{}. The detailed procedure is shown in algorithm \ref{algo}. The fixed bits allocated per sub-band are $\beta_x = [338, 237, 152, 135, 68, 10, 7, 3]$, designed to allocate more bits to the low frequency side. The FER threshold and additional bits are designed differently for low- and high-frequency sub-bands as $\lambda = [0.125, 0.125, 0.125, 0.125, 0.07, 0.07, 0.07, 0.07]$ and $\mathbf{\beta}_a = [7, 7, 5, 5, 3, 3, 3, 3]$, respectively. By using the proposed algorithm, the scale-factors for the sub-bands can be adaptively modified while considering the FER and frame gain calculated by the quantized LPC coefficients. 
\begin{algorithm}[t]
\caption{Adaptive target bits calculation}
\label{alg:adap_bits_cal}
\begin{algorithmic}
\REQUIRE $G(\nu)$ : frame gain, $\beta_x(b)$ : fixed bits for sub-bands, $\beta_a(b)$ : additional bits for sub-bands, $FER$ : frequency envelope ratio,  $\lambda(b)$ : thresholds of $FER$
\ENSURE $\beta_t(\nu,b)$ : number of target bits for sub-bands
\IF{$FER(\nu,b) \ge \lambda(b)- (G(\nu)>9.5 \,\,?\,\, 0.025\,\, :\,\, 0)$}
    \STATE $\bar{\beta}_a(b) = \beta_a(b)$
\ELSE 
    \STATE $\bar{\beta}_a(b) = 0$ 
\ENDIF
\STATE $\beta_t(\nu,b) = \beta_x(b) + \bar{\beta}_a(b)$
\RETURN $\beta_t(\nu,b)$
\end{algorithmic}
\label{algo}
\end{algorithm}
\subsection{Time domain aliasing augmentation}
\label{ssec:TDAA}

In the proposed framework, two audio coding modes are selectively operated depending on whether the CTNS are activated or not. However, the TDA caused by MDCT is not exactly canceled in the transition frames. In this section, a TDAA technique is proposed to tackle this problem. Before the elaboration of the proposed technique, the mathematical derivation for MDCT and its time-domain reconstruction are described. Consider the time-domain input signal $\mathbf{y}(\nu)$ as the column vector with length $N$: $\mathbf{y}(\nu)=[y(\nu N), y(\nu N+1), \ldots, y(\nu N+N-1)]^T.$ Further, the MDCT and inverse MDCT can be expressed as:
\begin{equation}
\begin{split}
\mathbf{Y}_{\mathrm{c}}(\nu)  =\mathbf{C} \mathbf{W}\begin{bmatrix}
\mathbf{y}^T(\nu-1) & \mathbf{y}^T(\nu)
\end{bmatrix}^T \\
{\begin{bmatrix}
\mathbf{y}_{\mathrm{c} 1}^T(\nu) & \mathbf{y}^T_{\mathrm{c} 2}(\nu)
\end{bmatrix}^T }  = \mathbf{C}^T \mathbf{Y}_{\mathrm{c}}(\nu),
\end{split}
\end{equation}
where the transformation matrix $\mathbf{C}\in\mathbb{R}^{N\times 2N}$ contains the MDCT basis functions as its rows, and the diagonal matrix $\mathbf{W}\in \mathbb{R}^{2N\times 2N}$ the identical analysis and synthesis window function as its diagonal elements. Although the reconstructed time-domain signals ($\mathbf{y}_{c1}$ and $\mathbf{y}_{c2}$) are contaminated by the TDA terms \cite{malvar1992signal}, the time-domain input signal $\mathbf{y}(\nu)$ can be perfectly reconstructed by the overlap and add (OLA) operation based on the TDA cancellation (TDAC) property of the MDCT. The OLA can be expressed as follows:
\begin{equation}
\mathbf{y}(\nu)=\mathbf{W}_2\mathbf{y}_{\mathrm{c} 2}(\nu)+\mathbf{W}_1\mathbf{y}_{\mathrm{c} 1}(\nu+1),
\end{equation}
where the $\mathbf{W}_1$ and $\mathbf{W}_2$ denote $N$-by-$N$ block matrices regarding the left and right side of the window function, respectively ($\mathbf{W} =blkdiag\left\{\mathbf{W}_1,\mathbf{W}_2\right\}$).

To decode an audio signal based on a proposed coding framework, the time-domain signal is reconstructed by the OLA procedure followed by a frame-based transform coding. When the adjacent coding modes are different, the coding artifacts occur owing to the uncompensated TDA terms. This is because the frame coded by the MDCT includes TDA terms, while the frame coded by the MCLT does not. To address this, we propose the TDAA algorithm. In the cases of the mode transition of the MDCT-frame followed by the MCLT-frame (case 1), and the MCLT-frame followed by MDCT-frame (case 2), the OLA procedures are modified as follows:
\begin{equation}
\mathbf{y}(\nu) = \begin{cases}
\mathbf{W}_2\mathbf{y}_{c2}(\nu)+\mathbf{W}_1\hat{\mathbf{y}}_1(\nu+1) &\text{(case 1)}\\
\mathbf{W}_{2}\hat{\mathbf{y}}_{2}(\nu)+\mathbf{W}_1\mathbf{y}_{c1}(\nu+1) &\text{(case 2)},
\end{cases}
\label{eq:case1}
\end{equation}
where $\hat{\mathbf{y}}_1$ and $\hat{\mathbf{y}}_2$ denote the left and right side of the augmented time-domain signal coded by the MCLT, respectively. Those augmented time-domain signals include artificially generated TDA terms made by multiplying the basis matrices ($\mathbf{C}^T\mathbf{C}$) to the reconstructed time-domain signal coded by the MCLT (${\mathbf{y}}_1$ and ${\mathbf{y}}_2$) as
\begin{equation}
\begin{bmatrix}
\hat{\mathbf{y}}^T_1(\nu) & \hat{\mathbf{y}}^T_2(\nu)
\end{bmatrix}^T = \mathbf{C}^T\mathbf{C} 
\begin{bmatrix}
\mathbf{y}^T_1(\nu) & \mathbf{y}^T_2(\nu)
\end{bmatrix}^T.
\end{equation}
The TDA terms in MDCT-frame are cancelled during the modified OLA (\ref{eq:case1}) with the augmented signal $\hat{\mathbf{y}}_1$ (case 1) or $\hat{\mathbf{y}}_2$ (case 2).

\begin{table}[t!]
\caption{Bit allocation for the proposed audio coding system}
\begin{minipage}[b]{1.0\linewidth}
  \centering
  \centerline{\includegraphics[width=5.5cm]{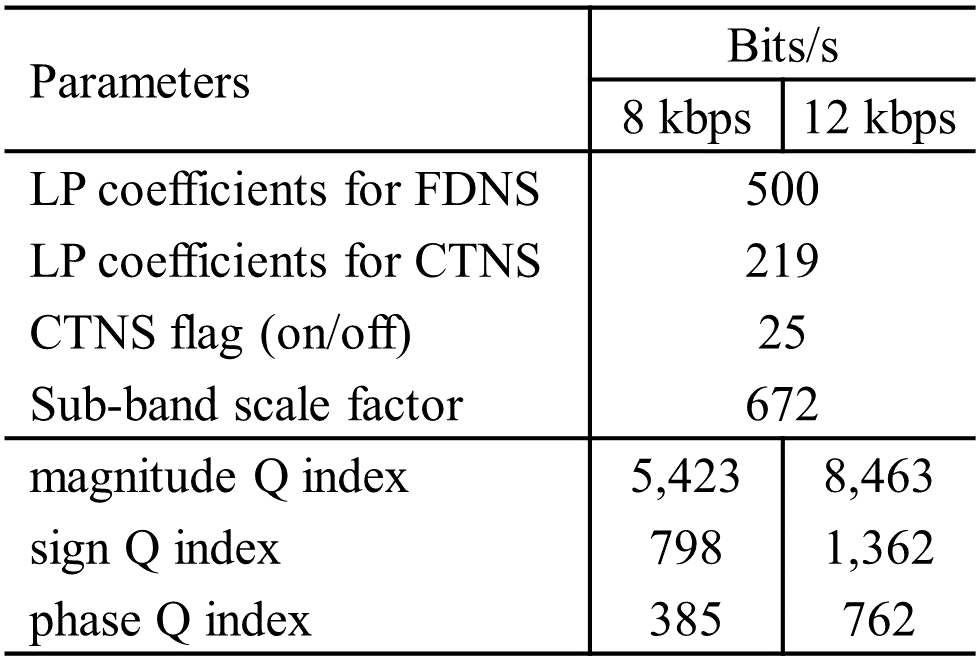}}
\vspace{-0.3cm}  
\end{minipage}
\label{tab:bit}
\end{table}

\begin{table}[t!]
\caption{Objective scores of TCX with USAC and the proposed system for bit rates of $8$ kbps and $12$ kbps}
\begin{minipage}[b]{1.0\linewidth}  
  \centering
  \centerline{\includegraphics[width=6.4cm]{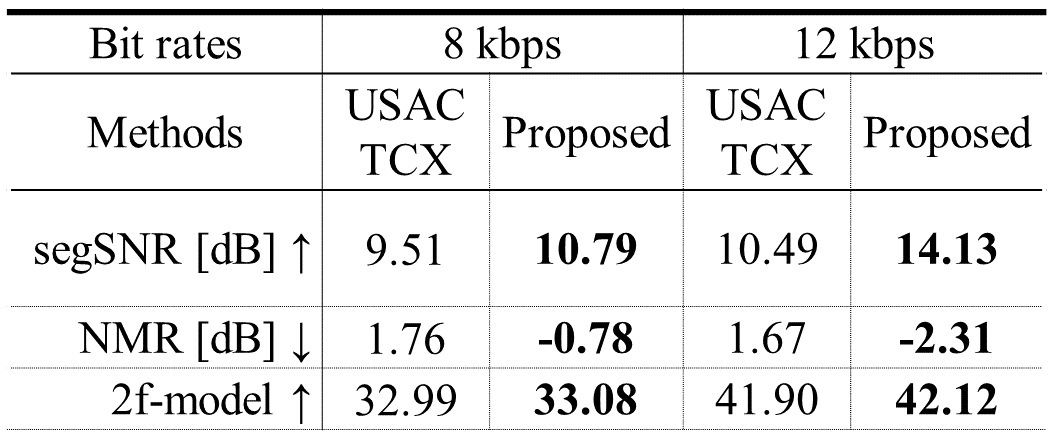}}  
\end{minipage}
\label{tab:objmea}
\vspace*{-3mm}
\end{table}

\section{Evaluation}
\label{sec:eval}
\subsection{Setup}
\label{ssec:setup}
To evaluate the proposed audio coding system, we calculated different objective metrics and conducted listening tests. Because our study aims to improve core-band coding performance at low bit rates, the target bit rates were chosen in the range of $8$ to $12$ kbps and the internal sampling rate was chosen as $12.8$ kHz. For a fair comparison with a single structure including the FDNS tool, long TCX in MPEG USAC \cite{usac} was adopted as the baseline method. In addition, since the segSNR in \cite{jo2023audio} at $12$ kbps ($11.83$ dB) was less than that of the proposed system, the system in \cite{jo2023audio} was not considered for further evaluations. To focus on the core-band coding artifacts of the systems, noise-filling and bandwidth-extension tools were not operated for both systems. The segSNR, noise-to-mask ratio (NMR), and 2f-model \cite{erlangen2020audiolabs,torcoli2021objective} were evaluated as objective metrics. The NMR and 2f-model were calculated based on the PEAQ function designed by McGill University \cite{kabal2002examination}. Subjective listening tests were conducted based on the MUSHRA methodology \cite{recommendation2001method} with the downsampled hidden reference and a $3.5$ kHz anchor. Each test was executed by $11$ experienced listeners who used high-quality listening headphones in a silence room. A total of $12$ test items were selected among the MPEG test sequences including $4$ speech, music, and mixed items.

The $1024$-sized sine window with $50\%$ overlap was utilized for MCLT analysis and synthesis. The LPC orders for the FDNS and CTNS were set to $16$ and $10$, respectively. The perceptual weights for the LP analyzes were set to $0.93$ and $0.94$, respectively. Here, the real-valued LPC coefficients are transformed to line spectral frequencies (LSFs) \cite{paliwal1992use} and quantized using $2$-stage vector quantizers (VQ) with $10$-bit codewords. Further, the magnitudes and phases of the complex roots for the complex-valued LP polynomials are directly quantized using the $3$-stage VQ with $11$-bit codewords. The allocated bits are described in Table \ref{tab:bit}. One bit for each frame is allocated for the switching flag. Similarly, one bit is assigned to the sign for a value with a non-zero magnitude. The bits for the scale factors were computed using the sample entropies of the quadruple scale factors (low and high bands). Analogously, the bits for the magnitudes and phases were calculated using the sample entropies of the quadruple frequency indices. Notably, the proper training for the entropy coder and its implementation are not considered because this study does not focus on developing a memory-efficient entropy coding tool.

\subsection{Discussions}
\label{ssec:result}
\begin{figure}[t]
\centering
\subfloat[]{
    \includegraphics[width=7.6cm]{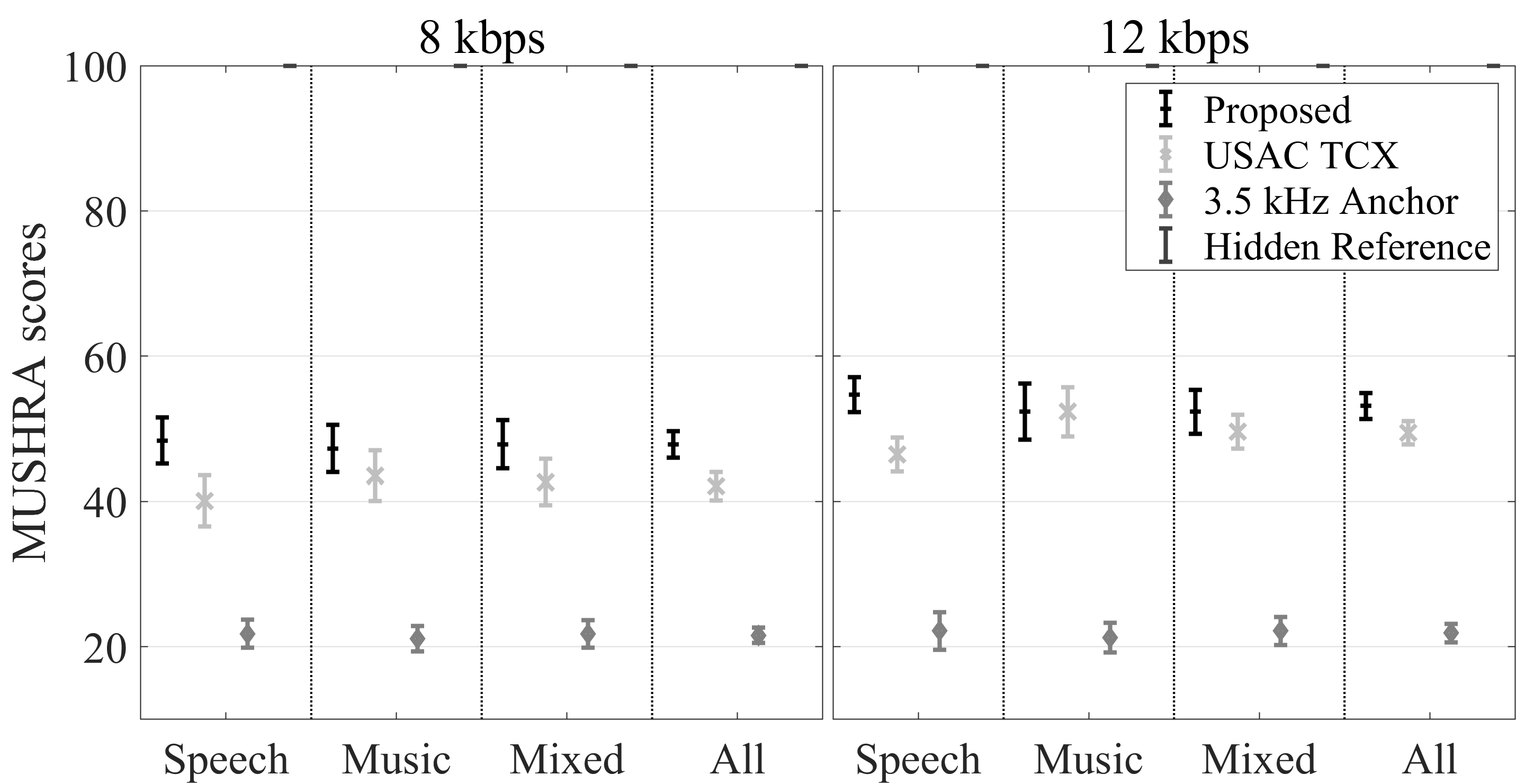}
    \label{fig:mushraAbs}
}
\hfill
\vspace{-0.05cm}
\subfloat[]{
    \includegraphics[width=7.6cm]{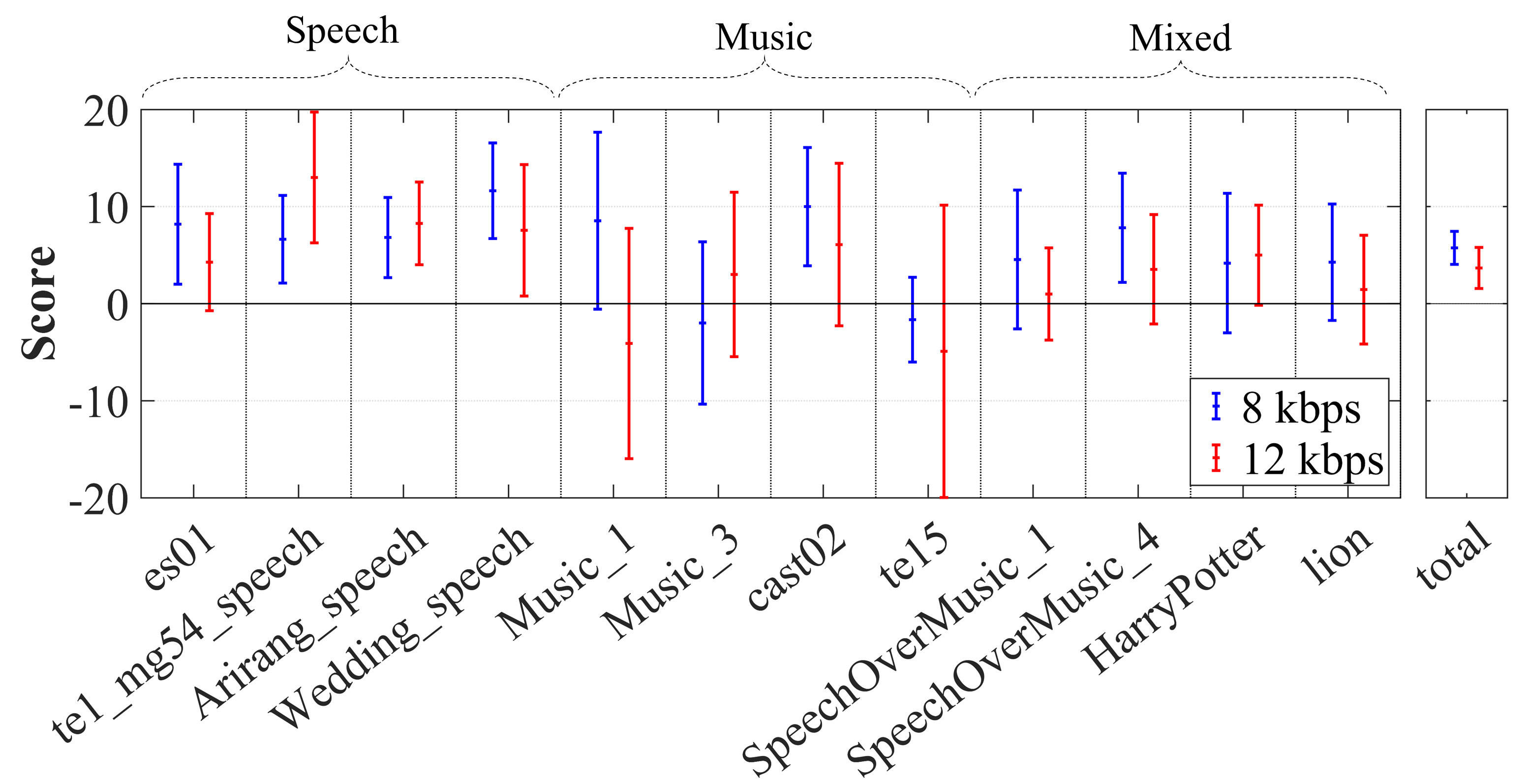}
    \label{fig:mushraDiff}
}
\caption{Absolute (a) and difference (b) MUSHRA scores and their $95$\% confidence intervals at $8$ kbps and $12$ kbps}
\vspace*{-3mm}
\end{figure}
The objective scores for both coding systems are summarized in Table \ref{tab:objmea}. All metrics of the proposed system are higher than those of the baseline system for two-bit rates. It should be noted that the segSNR and NMR values of the proposed system at $8$ kbps are higher than those of the baseline system at $12$ kbps. The 2f-model scores known to be highly correlated with subjective scores have marginal performance excellencies for both bit rates. In addition, to evaluate the effectiveness of TDAA in a separated manner, the segSNRs with and without TDAA were calculated at 12 kbps, yielding $14.13$ dB and $13.6$ dB, respectively. This shows that the proposed TDAA partially contributes the coding performance of the proposed framework. 

The absolute MUSHRA scores and their $95\%$ confidence intervals assuming a Student's t-distribution are illustrated in Fig. \ref{fig:mushraAbs}. The scores of the proposed system for the speech items are statistically higher than those of the baseline system for all the target bit rates. However, the scores for the mixed and music items are not statistically higher than those of the baseline system for all the bit rates although their mean values outperform those of the baseline system. To conclude, the total scores for all the items of the proposed system are statistically better than those of the baseline system for all the bit rates. To identify the variations in the performance between the different items, we computed the difference score for each item (Fig. \ref{fig:mushraDiff}). Except for `es01' at $12$ kbps, all the speech items have confidence intervals above zero. Contrarily, for the music category, relatively higher deviations are observed on a per-item basis (particularly for ``te15'', which has a high deviation and negative mean difference score). For the mixed category, all confidence intervals are greater than or equal to zero. In particular, the proposed system achieves statistically better audio quality for $6$ and $4$ out of $12$ items at $12$ and $8$ kbps, respectively.
\section{Conclusions}
\label{sec:conc}
We proposed a hybrid noise shaping integrated with the simplified FDNS and CTNS in the MCLT domain for low bit-rate audio coding. The proposed audio coding system includes adaptive target bit calculation using the frequency envelope information based on the quantized LPC coefficients and integrated quantizer for real and complex values to minimize quantization mismatches between different modes. Additionally, a time-domain aliasing augmentation algorithm artificially generates time-domain aliasing components during mode-switching operations. The scores of objective metrics and subjective listening evaluations demonstrate the superior performance of the proposed audio coding framework compared to the baseline system.

\bibliographystyle{IEEEtran}
\clearpage
\bibliography{refs.bib}

\begin{thebibliography}{10}
\providecommand{\url}[1]{#1}
\def\UrlFont{\rmfamily}
\providecommand{\newblock}{\relax}
\providecommand{\bibinfo}[2]{#2}
\providecommand\BIBentrySTDinterwordspacing{\spaceskip=0pt\relax}
\providecommand\BIBentryALTinterwordstretchfactor{4}
\providecommand\BIBentryALTinterwordspacing{\spaceskip=\fontdimen2\font plus
\BIBentryALTinterwordstretchfactor\fontdimen3\font minus
  \fontdimen4\font\relax}
\providecommand\BIBforeignlanguage[2]{{%
\expandafter\ifx\csname l@#1\endcsname\relax
\typeout{** WARNING: IEEEtran.bst: No hyphenation pattern has been}%
\typeout{** loaded for the language `#1'. Using the pattern for}%
\typeout{** the default language instead.}%
\else
\language=\csname l@#1\endcsname
\fi
#2}}

\bibitem{AMRWBp}
G.~T. 26.290, ``{Audio codec processing functions; Extended Adaptive Multi-Rate
  - Wideband (AMR-WB+) codec; Transcoding functions}.''

\bibitem{usac}
{ISO/IEC 23003-3}, ``{Information technology - MPEG audio technologies - Part
  3: Unified speech and audio coding},'' 2012.

\bibitem{evs}
G.~T. 26.441, ``{Codec for Enhanced Voice Services (EVS); General overview}.''

\bibitem{mpegh}
{ISO/IEC 23008-3}, ``{Information technology - High efficiency coding and media
  delivery in heterogeneous environments - Part 3: 3D audio - Amendment 1:
  MPEG-H, 3D audio profile and levels},'' 2015.

\bibitem{MPEG1}
{ISO/IEC 11172-3}, ``{Information technology - Coding of moving pictures and
  associated audio for digital storage media at up to about 1.5 Mbit/s - Part
  3: Audio},'' 1992.

\bibitem{AAC}
{{ISO/IEC} 14496}, ``{Information technology - Coding of Audiovisual Objects -
  Part 3, Audio},'' 1999.

\bibitem{bosi2002introduction}
M.~Bosi and R.~E. Goldberg, \emph{Introduction to digital audio coding and
  standards}.\hskip 1em plus 0.5em minus 0.4em\relax Springer Science \&
  Business Media, 2002, vol. 721.

\bibitem{gibson2005speech}
J.~D. Gibson, ``Speech coding methods, standards, and applications,''
  \emph{IEEE Circuits Syst. Mag.}, vol.~5, no.~4, pp. 30--49, 2005.

\bibitem{lefebvre1994high}
R.~Lefebvre, R.~Salami, C.~Laflamme, and J.-P. Adoul, ``{High quality coding of
  wideband audio signals using transform coded excitation (TCX)},'' in
  \emph{IEEE Int. Conf. Acoust. Speech Signal Process.}, vol.~1, 1994, pp.
  I--193.

\bibitem{xie1996embedded}
M.~Xie and J.-P. Adoul, ``{Embedded algebraic vector quantizers (EAVQ) with
  application to wideband speech coding},'' in \emph{IEEE Int. Conf. Acoust.
  Speech Signal Process.}, vol.~1, 1996, pp. 240--243.

\bibitem{ragot2004low}
S.~Ragot, B.~Bessette, and R.~Lefebvre, ``{Low-complexity multi-rate lattice
  vector quantization with application to wideband TCX speech coding at 32
  kbit/s},'' in \emph{IEEE Int. Conf. Acoust. Speech Signal Process.}, 2004.

\bibitem{fuchs2009mdct}
G.~Fuchs, M.~Multrus, M.~Neuendorf, and R.~Geiger, ``{MDCT-based coder for
  highly adaptive speech and audio coding},'' in \emph{Eur. Signal Process.
  Conf.}, 2009, pp. 1264--1268.

\bibitem{fuchs2015low}
G.~Fuchs, C.~R. Helmrich, G.~Markovi{\'c}, M.~Neusinger, E.~Ravelli, and
  T.~Moriya, ``{Low delay LPC and MDCT-based audio coding in the EVS codec},''
  in \emph{IEEE Int. Conf. Acoust. Speech Signal Process.}, 2015, pp.
  5723--5727.

\bibitem{vaalgamaa1999audio}
M.~Vaalgamaa, A.~H{\"a}rm{\"a}, and U.~K. Laine, ``Audio coding with auditory
  time-frequency noise shaping and irrelevancy reducing vector quantization,''
  in \emph{Audio Eng. Soc. Conf.}, 1999.

\bibitem{herre1996enhancing}
J.~Herre and J.~D. Johnston, ``{Enhancing the performance of perceptual audio
  coders by using temporal noise shaping (TNS)},'' in \emph{Audio Eng. Soc.
  Conv.}, 1996.

\bibitem{allamanche1999mpeg}
E.~Allamanche, R.~Geiger, J.~Herre, and T.~Sporer, ``{MPEG-4 low delay audio
  coding based on the AAC codec},'' in \emph{Audio Eng. Soc. Conv.}, 1999.

\bibitem{schnell2007enhanced}
M.~Schnell, R.~Geiger, M.~Schmidt, M.~Jander, M.~Multrus, G.~Schuller, and
  J.~Herre, ``{Enhanced MPEG-4 low delay AAC-Low bitrate high quality
  communication},'' in \emph{Audio Eng. Soc. Conv.}, vol. 6998, 2007.

\bibitem{jo2023audio}
B.~Jo, S.~Beack, and T.~Lee, ``{Audio coding with unified noise shaping and
  phase contrast control},'' in \emph{{IEEE Int. Conf. Acoust. Speech Signal
  Process.}}, 2023.

\bibitem{liu2008compression}
C.-M. Liu, H.-W. Hsu, and W.-C. Lee, ``Compression artifacts in perceptual
  audio coding,'' \emph{IEEE/ACM Trans. Audio, Speech, Lang. Process.},
  vol.~16, no.~4, pp. 681--695, 2008.

\bibitem{malvar1992signal}
H.~S. Malvar, ``Signal processing with lapped transforms,'' 1992.

\bibitem{malvar1999modulated}
H.~Malvar, ``A modulated complex lapped transform and its applications to audio
  processing,'' in \emph{IEEE Int. Conf. Acoust. Speech Signal Process.},
  vol.~3, 1999, pp. 1421--1424.

\bibitem{vafin2005entropy}
R.~Vafin and W.~B. Kleijn, ``Entropy-constrained polar quantization and its
  application to audio coding,'' \emph{IEEE Speech Audio Process.}, vol.~13,
  no.~2, pp. 220--232, 2005.

\bibitem{yoon2008coding}
B.-J. Yoon and H.~S. Malvar, ``{Coding overcomplete representations of audio
  using the MCLT},'' in \emph{Data Compress. Conf.}, 2008, pp. 152--161.

\bibitem{Jo2022modified}
B.~Jo, S.~Beack, and T.~Lee, ``Modified unrestricted polar quantization with
  the psychoacoustic parameter for audio coding,'' in \emph{Int. Congress.
  Acoustics}, 2022.

\bibitem{erlangen2020audiolabs}
A.~Erlangen, ``{AudioLabs-Subjective Evaluation of Blind Audio Source
  Separation Database: SEBASS-DB. Accessed: 2020-08-16},'' 2020.

\bibitem{torcoli2021objective}
M.~Torcoli, T.~Kastner, and J.~Herre, ``Objective measures of perceptual audio
  quality reviewed: An evaluation of their application domain dependence,''
  \emph{IEEE/ACM Trans. Audio, Speech, Lang. Process.}, vol.~29, pp.
  1530--1541, 2021.

\bibitem{kabal2002examination}
P.~Kabal, ``{An examination and interpretation of ITU-R BS. 1387: Perceptual
  evaluation of audio quality},'' \emph{TSP Lab Technical Report, McGill
  University}, pp. 1--89, 2002.

\bibitem{recommendation2001method}
I.-R. BS.1534, ``Method for the subjective assessment of intermediate sound
  quality ({MUSHRA}).''

\bibitem{paliwal1992use}
K.~K. Paliwal, ``On the use of line spectral frequency parameters for speech
  recognition,'' \emph{Digital signal process.}, vol.~2, no.~2, pp. 80--87,
  1992.

\end{thebibliography}
\end{sloppy}
\end{document}